\documentclass[aps,prl,reprint,showpacs,superscriptaddress]{revtex4-1}
\usepackage{graphicx} 
\usepackage{tabularx}
\usepackage{dcolumn} 
\usepackage{color}
\usepackage{bm}
\usepackage{amsmath}
\usepackage{amssymb}
\newcommand{\bi}[1]{\ensuremath{\boldsymbol{#1}}} 

\newcommand{\beq}{\begin{eqnarray}}
\newcommand{\eeq}{\end{eqnarray}}

\begin{document} 
 
\title{ 
UPt$_3$ as a Topological Crystalline Superconductor
}

\author{Yasumasa Tsutsumi} 
\affiliation{Condensed Matter Theory Laboratory, RIKEN, 
Wako, Saitama 351-0198, Japan} 
\author{Masaki Ishikawa} 
\affiliation{Department of Physics, Okayama University, 
Okayama 700-8530, Japan} 
\author{Takuto Kawakami} 
\affiliation{Department of Physics, Okayama University, 
Okayama 700-8530, Japan} 
\affiliation{International Center for Materials Nanoarchitectonics (WPI-MANA) National Institute for Materials Science, 
Tsukuba 305-0044, Japan} 
\author{Takeshi~Mizushima} 
\affiliation{Department of Physics, Okayama University, 
Okayama 700-8530, Japan} 
\author{Masatoshi Sato}
\affiliation{Department of Applied Physics, Nagoya University, 464-8603, Japan}
\author{Masanori Ichioka} 
\affiliation{Department of Physics, Okayama University, 
Okayama 700-8530, Japan} 
\author{Kazushige Machida} 
\affiliation{Department of Physics, Okayama University, 
Okayama 700-8530, Japan} 
\date{\today}

\begin{abstract} 
We investigate the topological aspect of the spin-triplet
 $f$-wave superconductor UPt$_3$ through microscopic calculations of
 edge- and vortex-bound states based on the quasiclassical Eilenberger and
 Bogoliubov-de Gennes theories. It is shown that a gapless and linear
 dispersion exists at the edge of the $ab$-plane. This forms a Majorana
 valley, protected by the mirror chiral symmetry. We also demonstrate that, 
 with increasing magnetic field,
 vortex-bound quasiparticles undergo a topological phase
 transition from topologically trivial states in the double-core vortex
 to zero-energy states in the normal-core vortex. As
 long as the ${\bm d}$-vector is locked into the $ab$-plane, the mirror
 symmetry holds the Majorana property of the zero-energy states, and thus
 UPt$_3$ preserves topological crystalline superconductivity that is robust
 against the crystal field and spin-orbit interaction. 
\end{abstract} 
 
 
 
\maketitle


{\it Introduction.---}
The unconventional aspect of the heavy-fermion superconductor UPt$_3$~\cite{stewart:1984} emerges as a multiple phase diagram in the temperature $T$ vs magnetic field $H$ plane, which is unique among a handful of strongly correlated superconductors.
In low fields, UPt$_3$ undergoes a double superconducting transition from a normal phase to the A-phase at $T_{\rm c1} \!\approx\! 550$ mK and from the A-phase to the B-phase at $T_{\rm c2} \!\approx\! 500$ mK~\cite{fisher:1989}.
The C-phase appears in the regime of low $T$'s and high $H$'s~\cite{adenwalla:1990, bruls:1990}. In spite of numerous works on UPt$_3$ over the past three decades following the discovery of superconductivity, the pairing mechanism and gap function have not been fully elucidated yet. 

A recent experiment has clarified the remarkable twofold symmetry breaking of the angle-resolved thermal conductivity in the $ab$-plane of the C-phase~\cite{machida:2012}. This convincingly suggests a spin-triplet $f$-wave function belonging to the $E_{1u}$ representation~\cite{tsutsumi:2012b}, where the gap function in the B-phase is described by the two-component ${\bm d}$-vector~\cite{machida:1999} and in the C-phase it reduces to a single component with the twofold symmetry breaking. Even within the B-phase, the ${\bm d}$-vector rotates from
$\bi{d}(\bi{k}) \!\propto \! \lambda_a\bi{b}+\lambda_b\bi{c}$ to $\lambda_a\bi{b}+\lambda_b\bi{a}$
with increasing ${\bm H}\!\parallel\! {\bm c}$ at the critical magnetic field $H_{\rm rot} \!\sim\! 2$ kG~\cite{tou:1996,tou:1998}, where 
$\lambda_{a,b}(\bi{k}) \!=\!k_{a,b}(5k_c^2-k^2)$
and $\bi{a}$, $\bi{b}$, and $\bi{c}$ are the unit vectors in a hexagonal crystal.
Most bulk thermodynamic experiments are understandable with the $E_{1u}$ scenario and another candidate based on the $E_{2u}$ representation~\cite{sauls:1994} described by $\bi{d}^{\prime } \!\propto\! \bi{c}(k_a+ik_b)^2k_c$ because both have point and line nodes in the B-phase~\cite{tsutsumi:2012b}.
The latter scenario
gives rise to the spontaneous breaking of the time-reversal symmetry in the B-phase and the fourfold symmetry breaking in the C-phase. These two scenarios differ in that the multicomponent order parameters originate from the multiple ${\bm d}$-vector in the $E_{1u}$ scenario and from the orbital degrees of freedom for the $E_{2u}$ representation.


In this Letter, we examine topological crystalline superconductivity in the B-phase of UPt$_3$ appropriate for the
$E_{1u}$ scenario with multiple ${\bm d}$-vectors. 
On the basis of a recent idea of Majorana fermions protected by crystal point group symmetries
\cite{ueno:2013,chiu:2013,zhang:2013},
it is demonstrated that the 
nontrivial topological property is directly linked to the orientation of
the ${\bm d}$-vector, and thus the field-induced rotation of the ${\bm
d}$-vector is accompanied by the topological phase transition of
vortex-bound states,
which is not observed in the $E_{2u}$ scenario.
Here, the topological aspects are unveiled through the
microscopic calculations of edge and vortex core states. It is shown
that zero-energy states exist at the edge of the $ab$-plane, which form
the topological ``Majorana valley''. Furthermore, employing numerical
calculations of the Bogoliubov-de Gennes (BdG) equation, we examine the
discretized quantum structure of quasiparticles (QPs) bound at a double-core vortex and a normal-core
vortex. 
It is found that increasing the magnetic field ${\bm H}\!\parallel\! {\bm c}$ 
induces a topological phase transition from 
topologically trivial states in the double-core vortex
to symmetry-protected Majorana fermions in a
normal-core vortex with ${\bm d}\!\perp\! {\bm c}$ via nontopological
Dirac fermions. The purposes of this Letter are to help identify the
pairing symmetry of UPt$_3$ and to place this material in the proper
position of topological crystalline superconductors. 



{\it Formulation.---}
Here, we utilize both the quasiclassical Eilenberger theory and the BdG theory. The former is valid for $\Delta \!\ll\! E_F$, which is well satisfied for most superconductors including UPt$_3$, where $\Delta$ and $E_{\rm F}$ denote the pair potential and Fermi energy, respectively.
The vortex-bound QP state is, however, discretized at $\Delta^2/E_{\rm F}$ intervals~\cite{caroli:1964}. The BdG theory enables us to describe the full quantum structure of low-lying QPs in the vortex state.

We start with the quasiclassical spinful Eilenberger theory~\cite{eilenberger:1968,serene:1983,sauls:2009,tsutsumi:2012b}.
The quasiclassical Green's function $\widehat{g}\!\equiv\! \widehat{g}(\bar{\bi{k}},\bi{r},\omega_n)$ is governed by the Eilenberger equation 
\begin{align}
-i\hbar\bi{v}(\bar{\bi{k}})\cdot\bi{\nabla }\widehat{g} 
= \left[
\begin{pmatrix}
i\hbar\omega_n\hat{1} & -\hat{\Delta }(\bar{\bi{k}},\bi{r}) \\
\hat{\Delta }^{\dagger }(\bar{\bi{k}},\bi{r}) & -i\hbar\omega_n\hat{1}
\end{pmatrix}
,\widehat{g}\right],
\label{Eilenberger eq}
\end{align}
with the normalization condition $\widehat{g}^2 \!=\! -\pi^2\widehat{1}$. The ordinary (wide) hat indicates the $2\!\times\!2$ ($4\!\times\! 4$) matrix in spin (particle-hole) space.
The quasiclassical Green's function is described in particle-hole space by
\begin{align}
\widehat{g}(\bar{\bi{k}},\bi{r},\omega_n) = -i\pi
\begin{pmatrix}
\hat{g}(\bar{\bi{k}},\bi{r},\omega_n) & i\hat{f}(\bar{\bi{k}},\bi{r},\omega_n) \\
-i\underline{\hat{f}}(\bar{\bi{k}},\bi{r},\omega_n) & -\underline{\hat{g}}(\bar{\bi{k}},\bi{r},\omega_n)
\end{pmatrix},
\end{align}
with the momentum on the Fermi surface $\bar{\bi{k}} \!=\! \bi{k}/k_{\rm F}\!=\! (k_a,k_b,k_c)/k_{\rm F}$, the center-of-mass coordinate $\bi{r}$, and the Matsubara frequency $\omega_n \!=\! (2n+1)\pi k_{\rm B} T/\hbar$ with $n \!\in\! \mathbb{Z}$.
The Fermi velocity is assumed as $\bi{v}(\bar{\bi{k}}) \!=\! v_{\rm F}\bar{\bi{k}}$ on a three-dimensional Fermi sphere.

The spin-triplet order parameter is expressed with the ${\bm d}$-vector as
$\hat{\Delta }(\bar{\bi{k}},\bi{r}) \!=\! i\bi{d}(\bar{\bi{k}},\bi{r})\cdot\hat{\bi{\sigma }}\hat{\sigma_b}$,
%
where $\hat{\bi{\sigma }}$ is the Pauli matrix.
The self-consistent condition for $\hat{\Delta }$ is given as
\begin{align}
\hat{\Delta}(\bar{\bi{k}},\bi{r}) \!=\! N_0\pi k_{\rm B}T\sum_{|\omega_n| \le \omega_c}\left\langle V(\bar{\bi{k}}, \bar{\bi{k}}') \hat{f}(\bar{\bi{k}}',\bi{r},\omega_n)\right\rangle_{\bar{\bi{k}}'},
\label{order parameter}
\end{align}
where $N_0$ is the density of states in the normal state.
The cutoff energy $\omega _{\rm c}$ is set to be
$\hbar\omega_{\rm c} \!=\! 20k_B T_{\rm c}$ with the transition temperature $T_{\rm c}$
and $\langle\cdots\rangle_{\bar{\bi{k}}}$ indicates the Fermi surface average.
In the B-phase without a magnetic field, the ${\bm d}$-vector is described by 
$\bi{d} \!=\! \Delta_1\lambda_a\bi{b}+\Delta_2\lambda_b\bi{c}$.
We neglect the splitting of $T_{\rm c}$ into $T_{\rm c1}$ and $T_{\rm c2}$ because the amplitudes of the two pair potentials, $\Delta _1$ and $\Delta _2$, are nearly equal at low temperatures in the B-phase.
The pairing interaction is
$V(\bar{\bi{k}}, \bar{\bi{k}}') \!=\!g[\lambda_a(\bar{\bi{k}})\lambda_a(\bar{\bi{k}}^{\prime})+\lambda_b(\bar{\bi{k}})\lambda_b(\bar{\bi{k}}^{\prime})]$, 
where the coupling constant $g$ is determined by
$(gN_0)^{-1}\!=\!\ln(T/T_c)+\pi k_BT\sum_{|\omega_n| \le \omega_{\rm c}}|\hbar\omega_n|^{-1}$.
We self-consistently solve Eqs.~\eqref{Eilenberger eq} and \eqref{order parameter} at $T \!=\! 0.5T_{\rm c}$.

By using the self-consistent solution of $\widehat{g}$ in Eqs.~(\ref{Eilenberger eq}) and (\ref{order parameter}), the spin current is calculated as
\begin{align}
\bi{j}_s^{\mu }(\bi{r})=\frac{\hbar }{2}N_0\pi k_{\rm B}T
\sum_{|\omega_n|\le\omega_c}\langle\bi{v}(\bar{\bi{k}}){\rm Im}[g_{\mu }(\bar{\bi{k}},\bi{r},\omega_n)]\rangle_{\bar{\bi{k}}},
\end{align}
where $g_{\mu }$ is defined as $\hat{g} \!=\! g_0 \hat{1} + {\bm g}\!\cdot\!\hat{{\bm \sigma}}$.
%
%
The local density of states (LDOS) for the energy $E$ is given by $N(\bi{r},E)\!=\! \langle N(\bar{\bi{k}},\bi{r},E)\rangle_{\bar{\bi{k}}}$, where the angle-resolved LDOS is 
\begin{align}
N(\bar{\bi{k}},\bi{r},E)
=N_0 {\rm Re} \left[g_0(\bar{\bi{k}},\bi{r}, \omega_n)|_{i\hbar\omega_n \rightarrow E+i\eta}\right] .
\end{align}
We here introduce a positive infinitesimal constant $\eta$, which is typically fixed at $\eta \!=\! 0.007\pi k_BT_c$.
To obtain $g_0(\bar{\bi{k}},\bi{r}, \omega_n)|_{i\hbar\omega_n \rightarrow E+i\eta}$, 
we solve Eq.~\eqref{Eilenberger eq} with $\eta -iE$ instead of $\hbar\omega_n$ under $\hat{\Delta}$ obtained self-consistently.

To obtain the discretized nature of vortex-bound states, we calculate the BdG equation. Since we here consider a straight vortex line along the $c$-axis (${\bm H} \!\parallel\! {\bm c}$), the wave number $k_c$ is a well-defined quantum number.
The BdG equation with a definite $k_c$ is given as~\cite{kaneko:2012}
%
\begin{multline}
\int d\bi{\rho }_2
\begin{pmatrix}
\hat{h}_{k_c}(\bi{\rho }_1,\bi{\rho }_2) & \hat{\Delta }_{k_c}(\bi{\rho }_1,\bi{\rho }_2) \\
-\hat{\Delta }_{-k_c}^{\dagger }(\bi{\rho }_1,\bi{\rho }_2) & -\hat{h}_{-k_c}(\bi{\rho }_1,\bi{\rho }_2)
\end{pmatrix}
\vec{u}_{\nu,k_c}(\bi{\rho }_2)\\
=E_{\nu,k_c}\vec{u}_{\nu,k_c}(\bi{\rho }_1),
\label{eq:BdG}
\end{multline}
where
$\hat{h}_{k_c}(\bi{\rho }_1,\bi{\rho }_2)  \!=\! \delta(\bi{\rho }_{12})(-\hbar^2\nabla_{\rm 2D}^2/2m-E_{\rm F}^{2{\rm D}}(k_c))\hat{1}$
with $\nabla_{\rm 2D}^2 \!=\! \partial_a^2+\partial_b^2$. The two-dimensional form of the Fermi energy $E_{\rm F}^{2{\rm D}}(k_c) \!=\! (\hbar^2/2m)(k_{\rm F}^2-k_c^2)$ reflects the $k_c$-cross section of the Fermi surface.
The order parameter in Eq.~(\ref{eq:BdG}) is obtained from the self-consistent solution of the quasiclassical theory and the relation
%
$\hat{\Delta }_{k_c}(\bi{\rho }_1,\bi{\rho }_2) \!=\!(2\pi)^{-2}\int d\bi{k}^{\rm 2D}\hat{\Delta }(\bi{k},\bi{\rho })e^{i\bi{k}^{\rm 2D}\cdot\bi{\rho }_{12}}$, 
%
where $\bi{\rho } \!=\! (\bi{\rho }_1+\bi{\rho }_2)/2$, $\bi{\rho }_{12} \!=\! \bi{\rho }_1-\bi{\rho }_2$, and $\bi{k}^{\rm 2D}$ are in the $ab$-plane.
Equation (\ref{eq:BdG}) describes QPs with the energy $E_{\nu,k_c}$ and the wave function $\vec{u}_{\nu,k_c} \!=\! ( u_{\nu,k_c}^{\uparrow }, u_{\nu,k_c}^{\downarrow }, v_{\nu,k_c}^{\uparrow }, v_{\nu,k_c}^{\downarrow })^{\rm T}$, where the index $\nu\in\mathbb{Z}$ denotes the $\nu$-th excited state of Eq.~(\ref{eq:BdG}).

{\it Edge states.---}
First, using the quasiclassical theory, we consider the edge state at the surface perpendicular to the $ab$-plane.
We here set a surface at $a \!=\! 0$ and impose the specular boundary condition on $\widehat{g}$ as $\widehat{g}(\bar{\bm k},{\bm r},\omega _n) \!=\! \widehat{g}(\bar{\bm k}_{\rm r},{\bm r},\omega _n)$ at $a \!=\! 0$, where $\bar{\bm k}_{\rm r}\!=\! \bar{\bm k} - 2{\bm a}\cdot\bar{\bm k}$.
In the B-phase without a magnetic field, the ${\bm d}$-vector is described by 
$\bi{d}(\bar{\bi{k}},a) \!=\! \Delta_{\perp }(a)\lambda_a(\bar{\bi{k}})\bi{b}+\Delta_{\parallel }(a)\lambda_b(\bar{\bi{k}})\bi{c}$.

\begin{figure}
\begin{center}
\includegraphics[width=8.5cm]{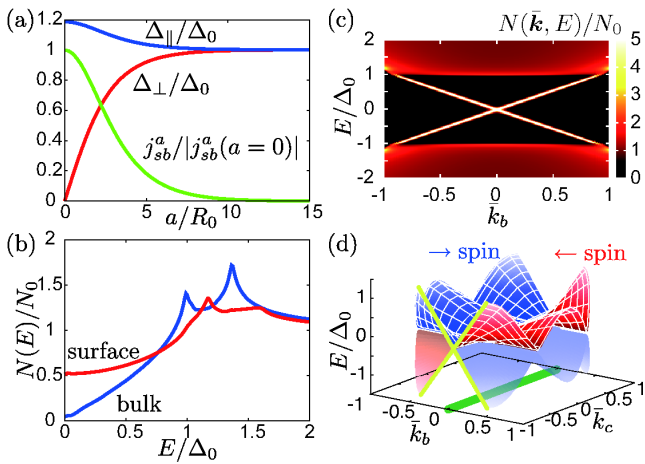}
\end{center}
\caption{\label{edge}(Color online) 
(a) Spatial profiles of order parameters and spin current $j_{sb}^a$ along the $a$-axis.
(b) LDOSs $N(E)$ at the surface and bulk.
(c) Angle-resolved LDOS $N(\bar{\bi{k}},E)$ at the surface as a function of $\bar{k}_b$ for $\bar{k}_c \!=\! 0$.
(d) Stereographic view of the dispersion of the surface bound state in $\rightarrow$- and $\leftarrow$-spin sectors.
}
\end{figure}

The spatial profile of the order parameter along the $a$-axis is shown in Fig.~\ref{edge}(a).
At $a\!=\! 0$, the specular boundary condition suppresses $\Delta_{\perp }$ coupled with a momentum $\bar{k}_a$ perpendicular to the surface.
In contrast, $\Delta_{\parallel}$ coupled to a parallel momentum $\bar{k}_b$ is enhanced by compensating for the loss of $\Delta_{\perp}$ at the surface.
Away from the surface, $\Delta_{\perp}$ increases and $\Delta_{\parallel }$ decreases toward the order parameter in the bulk B-phase $\Delta_0 \!=\! \Delta_{\perp } \!=\! \Delta_{\parallel }$. Figure~\ref{edge}(a) also shows the spin current $j_{sb}^{a}(a)$, implying that the $a$-component of the spin flows along the $b$-axis on the surface.

Figure~\ref{edge}(b) shows LDOS at the surface ($a \!=\! 0$) and bulk ($a/R_0\!=\! 40$), where $R_0=\hbar v_F/(2\pi k_{\rm B}T_{\rm c})$.
The two peaks at $E \!=\! \Delta_0$ and $E \!=\! (16\sqrt{15}/45)\Delta_0$ in the bulk LDOS are shifted to higher energies in the surface LDOS as a result of the enhancement of $\Delta_{\parallel}$ at the surface.
It is clearly seen that the zero-energy LDOS at the surface has
substantial weight (about half of the normal state), owing to the
dispersionless zero-energy state connecting
point nodes at the north and south poles of the Fermi sphere,
similarly to that in the superfluid $^3$He-A~\cite{tsutsumi:2011b}. 


We can separate the spin states by rotating the spin quantization axis to the $a$-axis as $\bi{d} \!=\! \Delta_0(\lambda_a\bi{b}+\lambda_b\bi{c}) \!=\!\Delta_0(\lambda_+\bi{d}_{\leftarrow }+\lambda_-\bi{d}_{\rightarrow })$, where $\lambda_{\pm } \!=\! \mp(\lambda_a\pm i\lambda_b)/\sqrt{2}$, $\bi{d}_{\rightarrow } \!=\! (\bi{b}+i\bi{c})/\sqrt{2}$, and $\bi{d}_{\leftarrow } \!=\! -(\bi{b}-i\bi{c})/\sqrt{2}$.
In Figs.~\ref{edge}(c) and \ref{edge}(d), we show
the angle-resolved LDOS as a function of $\bar{k}_b$ for $\bar{k}_c \!=\! 0$ and the dispersion of the surface bound state for the $\rightarrow$- and $\leftarrow$-spin states on the $\bar{k}_b\bar{k}_c$-plane.
The energy dispersion is derived from Eq.~(\ref{eq:BdG}) within the Andreev approximation as $E \!=\!\pm\Delta_0|5\bar{k}_c^2-1|\bar{k}_b$, where $+$ and $-$ correspond to the $\leftarrow$- and $\rightarrow$-spin sectors, respectively.
It is found that two linear branches with opposite slopes appear inside the bulk gap.
Since negative energy states are occupied at low $T$'s, the surface bound QPs in the $\rightarrow$- and $\leftarrow$-spin states counterflow along the $+{\bm b}$ and $-{\bm b}$ directions. Therefore, the spin current spontaneously appears along the edge of the $ab$-plane.
The dispersion of the surface bound state forms a ``Majorana valley" with a slope modulated along the $\bar{k}_c$ direction, as shown in Fig.~\ref{edge}(d). 
As is shown below, the Majorana valley in the B-phase is a direct consequence of the
topological crystalline superconductivity of UPt$_3$.



\begin{figure}
\begin{center}
\includegraphics[width=7.5cm]{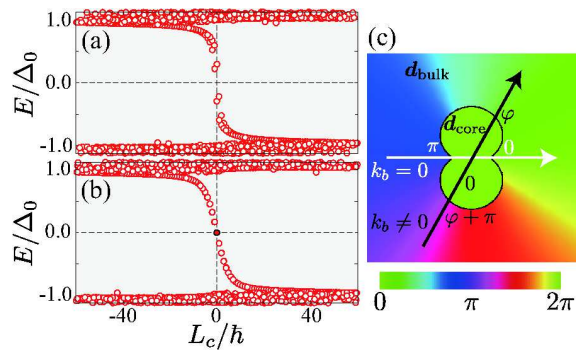}
\end{center}
\caption{\label{vortex}(Color online)
Energy spectra of QPs classified with the azimuthal angular momentum $L_c$ at $k_c\!= \!0$: (a) double-core vortex and (b) normal-core vortex. (c) Phase profiles of ${\bm d}_{\rm bulk}$ and ${\bm d}_{\rm core}$ and quasiclassical trajectories with $k_b\!=\!0$ (white arrow) and $k_b\!\ne\! 0$ (black arrow).}
\end{figure}

{\it Vortex states.---}
Next, using the BdG equation Eq.~\eqref{eq:BdG}, we clarify the vortex state under a magnetic field ${\bm H}\!\parallel\! {\bm c}$. The ${\bm d}$-vector is generalized to ${\bm d}({\bm k},{\bm \rho}) \!=\! {\bm d}_{\rm bulk}({\bm k},{\bm \rho}) + {\bm d}_{\rm core}({\bm k},{\bm \rho})$, where ${\bm d}_{\rm core}(|{\bm \rho}|\!\rightarrow\! \infty) \!=\! {\bm 0}$ denotes the component that fills the vortex core of ${\bm d}_{\rm bulk}$. In the low-$H$ regime of the B-phase, the bulk ${\bm d}$-vector is expressed as $\bi{d}_{\rm bulk} \!=\!\Delta_0e^{i\varphi }(\lambda_a\bi{b}+\lambda_b\bi{c})$ at $|{\bm \rho}|\!\rightarrow\! \infty$, where $\varphi$ is the azimuthal angle around the vortex core from the axis $a \!>\! 0$. 
The normal-core vortex
is characterized by ${\bm d}_{\rm core} \!=\! {\bm 0}$. 
The double-core vortex stabilized in the low-$H$ and low-$T$ regime has ${\bm d}_{\rm core} \!=\! \Delta_{\rm c}(\bi{\rho })\lambda_b\bi{a}$ where $\Delta_{\rm c}(\bi{\rho } \!=\! {\bm 0}) \!\neq\! 0$~\cite{tsutsumi:2012b}.
The momentum in ${\bm d}_{\rm core}$ is the same as that in the $c$-component of ${\bm d}_{\rm bulk}$ because the ${\bm d}$-vector easily rotates from the ${\bm c}$-axis to the ${\bm a}$-axis, as observed in the NMR Knight shift measurements~\cite{tou:1996,tou:1998}.
In the double-core vortex, as seen in Fig.~\ref{vortex}(c), 
the phases of ${\bm d}_{\rm bulk}(\varphi \!=\! 0)$ and ${\bm d}_{\rm core}$ are the same.


Figure \ref{vortex}(a) shows the energy spectrum of low-lying QPs in the double-core vortex with $k_c \!=\! 0$, obtained from the numerical diagonalization of Eq.~(\ref{eq:BdG}). All the eigenvalues are classified in terms of the angular momentum along the ${\bm c}$-axis, $L_c\!=\!-i\hbar\int d\bi{\rho }\vec{u}_{\nu,k_c}^{\dagger }(a\partial_b\!-\!b\partial_a)\vec{u}_{\nu,k_c}$.
It is seen from Fig.~\ref{vortex}(a) that zero-energy eigenstates are absent even in the vicinity of $L_c \!=\! 0$. 
To clarify the absence in the double-core vortex, let us consider the quasiclassical trajectories across the vortex core, as shown in Fig.~\ref{vortex}(c). The quasiclassical trajectory with the momentum $k_b \!=\! 0$ effectively feels the $\pi$-phase shift of the pair potential,
because the induced pair potential ${\bm d}_{\rm core} \!=\! \Delta_{\rm c}\lambda_b\bi{a}$ becomes zero for $k_b \!=\! 0$,
where the $\pi$-phase shift is necessary for the zero-energy state. In contrast, the trajectory with $k_b \!\neq\! 0$ feels ${\bm d}_{\rm core}$ interrupting the $\pi$-phase shift, which prevents the formation of the zero-energy state. Since the QP state at the vortex core is obtained as the superposition of all the contributions of the quasiclassical trajectories with various $k_b$'s, the zero-energy state is absent in the double-core vortex.

The normal-core vortex with ${\bm d}_{\rm core} \!=\! {\bm 0}$ is accompanied by the spin-degenerate zero-energy modes with $L_c \!=\! 0$, as shown in Fig.~\ref{vortex}(b). Within our model, the zero-energy states form the flat band along $k_c$. Note that, in a magnetic field $H \!\sim\! H_{\rm rot}$, the normal-core vortex lattice with a hexagonal symmetry is observed in the small-angle neutron scattering experiment~\cite{huxley:2000}. In the regime of $H \!<\! H_{\rm rot}$, the normal-core vortex is described by $\bi{d}(|\bi{\rho }|\rightarrow\infty) \!=\!\Delta_0e^{i\varphi }(\lambda_a\bi{b}+\lambda_b\bi{c})$. The zero-energy state is found to be fragile against the Zeeman field ${\bm H}\!\parallel\!{\bm c}$ and lifted to finite energies.
In the regime of $H\!>\!H_{\rm rot}$ where ${\bm d}(|\bi{\rho }|\rightarrow\infty) \!=\! \Delta_0e^{i\varphi }(\lambda_a\bi{b}+\lambda_b\bi{a})$, however, the zero-energy states with $L_c \!=\! 0$ remain robust against the magnetic field along the ${\bm c}$-axis,
because the $\uparrow$- and $\downarrow$-spin sectors of the ${\bm d}$-vector can be regarded as a spinless chiral superconductor.
Hence, in the normal-core vortex,
the excitation energy of the low-lying QP jumps to a zero-energy at the critical field
where the ${\bm d}$-vector is locked in the $ab$-plane. As described below, at $H \!=\! H_{\rm rot}$, the vortex-bound states undergo the topological phase transition associated with the mirror Chern number.


{\it Majorana fermions protected by mirror symmetries.---}
Finally, we clarify the symmetry protection and the Majorana nature of
zero-energy edge- and vortex-bound states in the B-phase of UPt$_3$. We
start with the BdG Hamiltonian,
\beq
\widehat{\mathcal{H}}({\bm k}) = 
\left( 
\begin{array}{cc}
\hat{\epsilon} ({\bm k}) & \hat{\Delta} ({\bm k}) \\ \hat{\Delta}^{\dag}({\bm k})
& - \hat{\epsilon}^{\rm T}(-{\bm k})
\end{array}
\right).
\eeq
Here, 
$\hat{\epsilon }({\bm k})$
is the Hamiltonian in the normal state of
UPt$_3$, which holds the D$_{\rm 6h}$ hexagonal symmetry. 
We find that two different mirror symmetries protect Majorana fermions in
the B-phase: One is the mirror reflection $\hat{\mathcal{M}}_{ca}$ with
respect to the $ca$-plane, which protects the Majorana valley on a
surface normal to the $a$-axis.
The other is the mirror reflection $\hat{\mathcal{M}}_{ab}$ with respect to the
$ab$-plane, which protects the Majorana zero mode in a vortex along the
$c$-axis. 
Below we show that the difference in symmetry gives rise to a
difference in Majorana nature between the edge- and
vortex-bound states. 
Note that UPt$_3$ shows an antiferromagnetic order in the normal state below about 5 K~\cite{heffner:1989}.
Even if the antiferromagnetic order coexists with the superconducting order,
the mirror symmetries are preserved macroscopically beyond the scale of the coherence length.

First, we consider the symmetry protection of the surface states.
Because the gap function in the B-phase is invariant
under the mirror reflection $\hat{\mathcal{M}}_{ca} \!\propto\! i\hat{\sigma }_b$, 
the BdG Hamiltonian $\widehat{\mathcal{H}}({\bm k})$ satisfies
$\widehat{\mathcal{M}}_{ca}\widehat{\mathcal{H}}({\bm
k})\widehat{\mathcal{M}}_{ca}^{\dagger} \!=\! \widehat{\mathcal{H}}(k_a,-k_b,k_c)$
with $\widehat{\mathcal{M}}_{ca} \!\equiv\! {\rm diag}(\hat{\mathcal{M}}_{ca},
\hat{\mathcal{M}}_{ca}^*)$. 
Therefore, combining the mirror symmetry with the time-reversal symmetry $\mathcal{T}$
and the particle hole symmetry $\mathcal{C}$, we have ``mirror  chiral
symmetry''
$\{\Gamma, 
\widehat{\mathcal{H}}({\bm k})\} \!=\! 0$ with $\Gamma \!=\! \mathcal{T}\mathcal{C}
\widehat{\mathcal{M}}_{ca}$ at $k_b \!=\! 0$~\cite{mizushima:2013}. 
The mirror chiral symmetry enables us to define the one-dimensional
winding number $w(k_c) \!=\! -(4\pi i)^{-1}\int_{-\pi}^{\pi}dk_a{\rm
tr}[\Gamma\widehat{\mathcal{H}}^{-1}\partial_{k_a}\widehat{\mathcal{H}}]$
\cite{sato:2009c,sato:2011}, which
is evaluated as $|w(k_c)| \!=\! 2$ for $k_b \!=\! 0$, $|k_c| \!<\! k_{\rm F}$ and $w(k_c) \!=\! 0$ for
other $k_b$'s and $k_c$'s. 
Thus, the system is topologically non-trivial and the bulk-edge
correspondence ensures the existence of the Majorana valley in
Fig.~\ref{edge}(d) with a flat dispersion connecting the point nodes as
$E \!=\! 0$ at $k_b \!=\! 0$ and $|k_c| \!<\! k_{\rm F}$. 
In addition, owing to the mirror chiral symmetry, the Majorana valley shows the
Majorana Ising anisotropy that the surface bound states are gapped only
by a magnetic field along the $b$-axis \cite{mizushima:2012b}.
A magnetic field in the $ca$-plane or the ${\bm d}$-vector
rotation in the high-field phase in the B-phase does not obscure the
topological protection since the combination of the mirror reflection
$\hat{\mathcal {M}}_{ca}$ and the time-reversal is not broken, but each
of them is broken.
Here, note that, while the Majorana valley has a close similarity to the
topological Fermi arcs in $^3$He-A~\cite{volovik:2013,silaev:2012}, the arcs' topological origins are
totally different: 
The time-reversal breaking is essential for the topological Fermi arcs,
but not for the Majorana valley. 

For the topological protection of zero-energy states in a vortex, the
mirror reflection $\hat{\mathcal{M}}_{ab} \!\propto\! i\hat{\sigma}_c$ with
respect to the $ab$-plane is essential.
Following Ref.~\onlinecite{ueno:2013}, one can show that, if the gap function is odd under the mirror reflection $\hat{\mathcal{M}}_{ab}$, $\hat{\mathcal{M}}_{ab} \hat{\Delta }({\bm k})\hat{\mathcal{M}}_{ab}^{\rm T} \!=\! -\hat{\Delta }(k_a, k_b, -k_c)$, a normal-core vortex may support the Majorana zero mode protected by the mirror symmetry:
In this case, $\widehat{\mathcal{H}}({\bm k})$ commutes with the mirror operator $\widehat{\mathcal{M}}_{ab}^{(-)} \!\equiv\! {\rm diag}(\hat{\mathcal{M}}_{ab},-\hat{\mathcal{M}}_{ab}^*)$.
On the mirror reflection invariant plane $k_c \!=\! 0$, the system splits into two subsectors with two different eigenvalues of $\widehat{\mathcal{M}}_{ab}^{(-)}$, and because of the minus sign in front of $\hat{\mathcal{M}}_{ab}^*$ in the mirror operator $\widehat{\mathcal{M}}_{ab}^{(-)}$, each mirror subsector supports its own particle-hole symmetry.
This means that 
the mirror subsectors are topologically equivalent to class D of the table in
Ref.~\onlinecite{schnyder:2008} and thus the nontrivial Chern number in each
subsector ensures non-Abelian Majorana fermions~\cite{ueno:2013,sato:cond1305}.




Since 
${\bm d}({\bm k}) \!\propto\! \lambda_a{\bm b} + \lambda_b{\bm c}$
does not have a definite mirror parity under
$\hat{\mathcal{M}}_{ab}$, the spin-orbit interaction or crystal
field, which is ignored in 
the numerical calculations above,
lifts zero-energy states, implying that the B-phase with the configuration of
such a ${\bm d}$-vector is topologically trivial for vortex-bound states. 
On the other hand, for 
${\bm d}({\bm k}) \!\propto \! \lambda_a{\bm b} + \lambda_b{\bm a}$
rotated by a high magnetic field ${\bm
H}\!\parallel\! {\bm c}$, the gap function is odd under
the mirror reflection $\hat{\mathcal{M}}_{ab}$, and Majorana vortex-bound states
protected by the mirror symmetry are possible. 
Actually, for UPt$_3$ with five closed Fermi
surfaces~\cite{mcmullan:2008}, the parity of the mirror Chern number at $k_c \!=\! 0$ is
odd~\cite{sato:2010}. This ensures that 
there exist Majorana zero modes in a
vortex along the $c$-axis. Hence, the low-lying QPs bound at the
normal-core vortex undergo the topological phase transition from
nontopological zero modes to symmetry protected Majorana fermions with
increasing magnetic field. 
The topological phase transition without closing the bulk gap but accompanied by symmetry breaking has also been discussed in Refs.~\onlinecite{mizushima:2012b} and \onlinecite{ezawa:2013} recently.



{\it Concluding remarks.---}
We have investigated the topological aspect of edge- and vortex-bound
states for the recently identified gap function of the UPt$_3$ B-phase. 
In the edge state, Majorana fermions with linear dispersion are bound
and their zero-energy states form the Majorana valley. 
The Majorana valley 
is protected by the mirror chiral symmetry, responsible for Ising anisotropy.


Note that the symmetry-protected Majorana valley at the surface can be detected by
tunneling spectroscopy~\cite{tanaka:1995,kashiwaya:2000}:
The flat dispersion gives rise to a finite zero bias tunneling conductance, 
where the tunneling conductance is related to the surface LDOS [Fig.~\ref{edge}(b)] in the low transparent limit.
The Majorana Ising anisotropy results in a decrease in the zero bias conductance under a magnetic field only along the $b$-axis.
In contrast, the surface states in the $E_{2u}$ scenario
are not coupled with a magnetic field along the $b$-axis.

We have also demonstrated that the double-core vortex is not accompanied by the
zero-energy state. As $H$
increases, the finite energy excitations in a double-core vortex undergo
the topological transition to symmetry-protected Majorana fermions via
topologically trivial zero modes in a normal-core vortex. The Majorana
fermions are protected by a mirror symmetry against perturbations, such
as a magnetic field, a crystal field, and a spin-orbit interaction, when the
${\bm d}$-vector is locked in the $ab$-plane,
${\bm d}({\bm k}) \!\propto\! \lambda_a\bi{b}+\lambda_b\bi{a}$.
Hence, the B-phase of UPt$_3$
offers a promising platform for studying topological crystalline
superconductors. 

Some of the numerical calculations were performed using the RIKEN Integrated Cluster of Clusters (RICC).
This work was supported by KAKENHI (Nos. 24840048, 21340103, 22103005,
2200247703, 25287085, and 25103716). 


\begin{thebibliography}{10}

\bibitem{stewart:1984}
G.~R. Stewart, Z.~Fisk, J.~O. Willis, and J.~L. Smith: Phys. Rev. Lett.
  {\bfseries 52} (1984) 679.

\bibitem{fisher:1989}
R.~A. Fisher, S.~Kim, B.~F. Woodfield, N.~E. Phillips, L.~Taillefer,
  K.~Hasselbach, J.~Flouquet, A.~L. Giorgi, and J.~L. Smith: Phys. Rev. Lett.
  {\bfseries 62} (1989) 1411.

\bibitem{adenwalla:1990}
S.~Adenwalla, S.~W. Lin, Q.~Z. Ran, Z.~Zhao, J.~B. Ketterson, J.~A. Sauls,
  L.~Taillefer, D.~G. Hinks, M.~Levy, and B.~K. Sarma: Phys. Rev. Lett.
  {\bfseries 65} (1990) 2298.

\bibitem{bruls:1990}
G.~Bruls, D.~Weber, B.~Wolf, P.~Thalmeier, B.~L\"uthi, A.~d. Visser, and
  A.~Menovsky: Phys. Rev. Lett. {\bfseries 65} (1990) 2294.

\bibitem{machida:2012}
Y.~Machida, A.~Itoh, Y.~So, K.~Izawa, Y.~Haga, E.~Yamamoto, N.~Kimura,
  Y.~{\=O}nuki, Y.~Tsutsumi, and K.~Machida: Phys. Rev. Lett. {\bfseries 108}
  (2012) 157002.

\bibitem{tsutsumi:2012b}
Y.~Tsutsumi, K.~Machida, T.~Ohmi, and M.~Ozaki: J. Phys. Soc. Jpn. {\bfseries
  81} (2012) 074717.

\bibitem{machida:1999}
K.~Machida, T.~Nishira, and T.~Ohmi: J. Phys. Soc. Jpn. {\bfseries 68} (1999)
  3364.

\bibitem{tou:1996}
H.~Tou, Y.~Kitaoka, K.~Asayama, N.~Kimura, Y.~{\=O}nuki, E.~Yamamoto, and
  K.~Maezawa: Phys. Rev. Lett. {\bfseries 77} (1996) 1374.

\bibitem{tou:1998}
H.~Tou, Y.~Kitaoka, K.~Ishida, K.~Asayama, N.~Kimura, Y.~{\=O}nuki,
  E.~Yamamoto, Y.~Haga, and K.~Maezawa: Phys. Rev. Lett. {\bfseries 80} (1998)
  3129.

\bibitem{sauls:1994}
J.~A. Sauls: Adv. Phys. {\bfseries 43} (1994) 113.

\bibitem{ueno:2013}
Y.~Ueno, A.~Yamakage, Y.~Tanaka, and M.~Sato: Phys. Rev. Lett. {\bfseries 111}
  (2013) 087002.
  \label{ueno}

\bibitem{chiu:2013}
C.-K. Chiu, H.~Yao, and S.~Ryu: Phys. Rev. B {\bfseries 88} (2013) 075142.

\bibitem{zhang:2013}
F.~Zhang, C.~L. Kane, and E.~J. Mele: Phys. Rev. Lett. {\bfseries 111} (2013)
  056403.

\bibitem{caroli:1964}
C.~Caroli, P.~de~Gennes, and J.~Matricon: Phys. Lett. {\bfseries 9} (1964) 307.

\bibitem{eilenberger:1968}
G.~Eilenberger: Z. Phys. {\bfseries 214} (1968) 195.

\bibitem{serene:1983}
J.~W. Serene and D.~Rainer: Phys. Rep. {\bfseries 101} (1983) 221.

\bibitem{sauls:2009}
J.~A. Sauls and M.~Eschrig: New J. Phys. {\bfseries 11} (2009) 075008.

\bibitem{kaneko:2012}
S.~Kaneko, K.~Matsuba, M.~Hafiz, K.~Yamasaki, E.~Kakizaki, N.~Nishida,
  H.~Takeya, K.~Hirata, T.~Kawakami, T.~Mizushima, and K.~Machida: J. Phys.
  Soc. Jpn. {\bfseries 81} (2012) 063701.

\bibitem{tsutsumi:2011b}
Y.~Tsutsumi, M.~Ichioka, and K.~Machida: Phys. Rev. B {\bfseries 83} (2011)
  094510.

\bibitem{huxley:2000}
A.~Huxley, P.~Rodi{\`e}re, D.~M. Paul, N.~van Dijk, R.~Cubitt, and J.~Flouquet:
  Nature {\bfseries 406} (2000) 160.

\bibitem{heffner:1989}
R.~H. Heffner, D.~W. Cooke, A.~L. Giorgi, R.~L. Hutson, M.~E. Schillaci, H.~D.
  Rempp, J.~L. Smith, J.~O. Willis, D.~E. MacLaughlin, C.~Boekema, R.~L.
  Lichti, J.~Oostens, and A.~B. Denison: Phys. Rev. B {\bfseries 39} (1989)
  11345.

\bibitem{mizushima:2013}
T.~Mizushima and M.~Sato: New J. Phys. {\bfseries 15} (2013) 075010.

\bibitem{sato:2009c}
M.~Sato and S.~Fujimoto: Phys. Rev. B {\bfseries 79} (2009) 094504.

\bibitem{sato:2011}
M.~Sato, Y.~Tanaka, K.~Yada, and T.~Yokoyama: Phys. Rev. B {\bfseries 83}
  (2011) 224511.

\bibitem{mizushima:2012b}
T.~Mizushima, M.~Sato, and K.~Machida: Phys. Rev. Lett. {\bfseries 109} (2012)
  165301.
  \label{mizushima}

\bibitem{volovik:2013}
G.~E. Volovik: J. Supercond. Nov. Magn. {\bfseries 26} (2013) 2887.

\bibitem{silaev:2012}
M.~A. Silaev and G.~E. Volovik: Phys. Rev. B {\bfseries 86} (2012) 214511.

\bibitem{schnyder:2008}
A.~P. Schnyder, S.~Ryu, A.~Furusaki, and A.~W.~W. Ludwig: Phys. Rev. B
  {\bfseries 78} (2008) 195125.
  \label{schnyder}

\bibitem{sato:cond1305}
M.~Sato, A.~Yamakage, and T.~Mizushima: to be published in Physica E; arXiv:1305.7469.

\bibitem{mcmullan:2008}
G.~J. McMullan, P.~M.~C. Rourke, M.~R. Norman, A.~D. Huxley, N.~Doiron-Leyraud,
  J.~Flouquet, G.~G. Lonzarich, A.~McCollam, and S.~R. Julian: New J. Phys.
  {\bfseries 10} (2008) 053029.

\bibitem{sato:2010}
M.~Sato: Phys. Rev. B {\bfseries 81} (2010) 220504.

\bibitem{ezawa:2013}
M.~Ezawa, Y.~Tanaka, and N.~Nagaosa: Sci. Rep. {\bfseries 3} (2013) 2790.
\label{ezawa}

\bibitem{tanaka:1995}
Y.~Tanaka and S.~Kashiwaya: Phys. Rev. Lett. {\bfseries 74} (1995) 3451.

\bibitem{kashiwaya:2000}
S.~Kashiwaya and Y.~Tanaka: Rep. Prog. Phys. {\bfseries 63} (2000) 1641.

\end{thebibliography}

\end{document}